\documentclass[12pt,halfline,a4paper]{ouparticle}
\usepackage{amssymb,amsfonts,amsthm,amsmath}
\usepackage{mathrsfs}

\begin{document}

\title{FRACTIONAL CALCULUS AND THE ESR TEST}

\author{%
\name{J. Vanterler da C. Sousa and E. Capelas de Oliveira}
\address{Department of Applied Mathematics - Imecc,\\
Unicamp, 13083-859, Campinas, SP, Brazil}
\email{ra160908@ime.unicamp.br, capelas@ime.unicamp.br}
\and
\name{L. A. Magna}
\address{Department of Medical Genetics, School of Medical Sciences,\\
Unicamp, 13083-887, Campinas, SP, Brazil}
\email{amagna@uol.com.br}}

\abstract{We consider a partial differential equation
associated with a mathematical model describing the concentration of
nutrients in blood which interferes directly on the erythrocyte
sedimentation rate in the case of an average fluid velocity equal to zero.
Introducing the fractional derivative in the Caputo sense, we propose a
time-fractional mathematical model which contains, as a particular case, the
model proposed by Sharma et al.\ \cite{GMR}. Our main purpose is to obtain an
analytic solution of this time-fractional partial differential equation in terms of
the Mittag-Leffler function and Wright function.}

\date{\today}


\maketitle

\section{Introduction} 

In 1897, the Polish physician E. F. Biernacki introduced a blood test that
helped in diagnosing the acute phase of inflammatory diseases and in 
following up body inflammation itself, known as Erythrocyte
Sedimentation Rate (ESR) \cite{EKU,EJK,ASJJ}.  The discovery was announced in
two articles: the first, in Polish, in Lekarska Gazeta, the second, in German,
in Deutsche Medizinische Wochenschrift \cite{EB,EB1}.  On the other hand,
at the beginning of nineteenth century, Robin Fahraeus and A. Westergren, when
performing pregnancy and tuberculosis tests, developed a test similar to 
ESR known as the Fahraeus-Westergren test  \cite{AW,AW1,RF,RF1}.

Nowadays, due to the discovery of new and more accurate methods, ESR is little
used despite it being a quick and low cost test.  Nevertheless, the test is
still recommended for patients with suspected giant cell arthritis, rheumatics
polymyalgia and rheumatoid arthritis, among others \cite{MBR}. However, as ESR
is not quite specific, it is necessary to conduct further tests to confirm the
result obtained by ESR in order to avoid false-positive and false-negative
results, which are likely to occur in the presence of some factor whose
influence on blood properties affects the test results \cite{NEL,JDA,SEB}, e.g.:

\begin{itemize}

\item Analytic factors such as an inclined tube and ambient temperature, which
	would respectively increase and decrease ESR \cite{MMFD}. Other factors
	which affect the results are the presence of external vibration and
	tube deformation \cite{PSRD}.

\item Physiological and pathological factors such as anemia due to low
	erythrocyte concentration, pregnancy and old age, 
	resulting in increased ESR; polycythemia and increased leukocyte
	counting, resulting in decreased ESR. 

\end{itemize}

The concentration of nutrients in blood also plays a role in the analysis of
ESR results \cite{IPHJ}. Moreover, Nayha \cite{SNA} noted that people who drink
coffee and smoke present higher values of ESR. The use of some types of
anticoagulants such as Sodium citrate, oxalate or $K_{3}$ EDTA is also
responsible for influencing the test results \cite{ICH,ICH1,KLBA,BSG}.

Huang et al.\ \cite{JCPA} published a work in which they measured the
concentration of red cells in blood at different times in samples of 5 male
people. In the same year, Huang et al.\ \cite{CJHC} developed a mathematical
model to describe the behavior of the concentration of blood cells, giving
importance to mobile boundary problems. Another notable work in ESR context was
written by Sartory \cite{SAR}, whose objective was to study the prediction of
erythrocyte sedimentation profiles.  Moved by Huang's 1971 work, in 1990 Reuben
and Shannon \cite{RS} discussed some problems in the mathematical modeling of
concentration of red blood cells. However, the authors of those studies did not
take into account the transfer of nutrients from capillaries to tissues.  Due
to this fact, Sharma et al.\ \cite{GMR} established a mathematical model taking
into account the transfer of nutrients, making it a more precise model.

Thus the ESR test can be used to obtain several clinical diagnoses and may be
studied as a particular type of transport phenomena \cite{EL}.  It is worth
mentioning that there are several transport phenomena whose fractional versions provide better descriptions than the corresponding classical models
\cite{ECR,FG,RJ}.

Those models are generally constructed with the help of nonlinear partial
differential equations whose solutions require numerical methods to be
discussed. On the other hand, the corresponding linear partial differential
equations can be solved by means of some analytic method, but the solutions
sometimes do not actually describe a particular phenomenon.

The main motivation for the study of concentration $C(x,t)$ by means of
fractional calculus is the discrete mathematical model proposed by Sharma et al. \cite{GMR}. In this article we present an application of fractional calculus,
specifically of fractional derivatives in the Caputo sense, to study the
profile of $C(x,t)$. Event though we have the same purpose as Sharma et al. \cite{GMR}, the equivalence, in an appropriate limit, between our results and
the integer order case is not immediate, because the way they develop their
calculations with integer order derivatives has different directions from the
calculations using fractional derivatives in the Caputo sense, and each step
has to be checked.

As we are seeking the same goal, it is useful to make some comparisons in order
to highlight the developments brought by this work. Indeed, it is not an easy
task, and much less trivial, to find the analytic solution of the proposed
fractional model and to present its behavior graphically.  The main differences
and similarities with regard to the integer case are:

\begin{itemize}

\item We assume an average speed $U = 0$, as in Sharma et al. model \cite{GMR},
	thus restricting ourselves to the diffusion model.  We use this model
		to introduce the basic concepts of fractional calculus and to
		present our fractional mathematical model.

\item We propose a model with fractional derivatives in the Caputo sense with a
	time derivative of order $ 0 <\mu\leq 1$. Consequently, the solution is
	dependent on the parameter $\mu$. In the limit $\mu\rightarrow 1$ we
	recover the solution of the Sharma et al.\ \cite{GMR} as a particular
	case.

\item What is expected is that the solutions of each model are distinct, and 
	that is what actually happens. Indeed, the analytic solution
	obtained by Sharma et al.\ \cite{GMR} is expressed as a product of the
	exponential function $\mbox{exp} (\cdot{})$ and the complementary error
	function $\mbox{erfc}(\cdot)$.  On the other hand, the solution
	obtained for the fractional mathematical model is given in terms of the
	Mittag-Leffler function and the Wright function.

\item The analytic behavior of the solution obtained with fractional
	derivatives allows a more detailed analysis, because we have an extra
	degree of freedom in parameter $\mu$ ($0<\mu \leq $ 1), which
	permits better fitting of experimental data on nutrient
	concentration in blood.

\end{itemize}

We should point out that the methodology to discuss a problem involving a
particular fractional derivative is growing in all fields of the knowledge,
specifically the integral transform methodology. 

This paper is organized as follows: In section two we introduce the so-called
fractional mathematical model associated with ESR, a generalization of the
model proposed by Sharma et al.\ \cite{GMR}, which will be recovered through a
convenient limit process. Section three, our main result, is dedicated to
obtain the analytic solution of our model, which is found using the methodology
of Laplace transform and is expressed in terms of the Mittag-Leffler function
and the Wright function. We also present a graphical analysis of the solution.
In section four we recover as a special case, through a convenient limit
process, the solution found by Sharma et al.\ \cite{GMR}.


\section{Time-fractional partial differential equation }

In this section we present a fractional version of the linear partial
differential equation (PDE) associated with the mathematical model
of Sharma et al.\ \cite{GMR}  used to describe the behavior of the concentration of nutrients in
blood cells, a factor that directly affects ESR. We assume that the average
fluid velocity is equal to zero. Our model can be considered a generalization of
the Sharma et al. \cite{GMR} model, in the sense that it recovers the latter as
a special case, as we shall see in section four.

The blood concentration of nutrients $C(x,t)$ satisfies the following
non-homogeneous time-fractional PDE,

\begin{equation}  \label{S1}
D_{L}\mathcal{D}_{x}^{2}C\left( x,t\right) - \mathcal{D}_{t}^{\mu }C\left(
x,t\right) =\phi \left( x,t\right) ,
\end{equation}
with $0<\mu\leq 1$, where $D_{L}$ is a positive constant and $\phi(x,t)$ is a function
describing the nutrient transfer rate and which satisfies the 
partial differential equation:

\begin{equation}  \label{S2}
D\mathcal{D}_{x}^{2}\phi \left( x,t\right) -k\phi \left( x,t\right) -%
\mathcal{D}_{t}\phi \left( x,t\right) =0,
\end{equation}
with $D$ and $k$ both positive constants.

The initial and boundary conditions imposed here are given by

\begin{equation*}
\left\{ 
\begin{array}{ll}
\phi (x,0)=\exp \left( -\sqrt{\frac{k-a}{D}}x\right) , & k\geq a,D>0 , \\ 
\phi (0,t)=\exp \left( -at\right) , & t>0 , \\ 
\phi (\infty ,t)=0, & t>0.
\end{array}%
\right.
\end{equation*}

The solutions of the Eq.(\ref{S2}) can be written as

\begin{equation*}
\phi \left( x,t\right) =\exp \left( -\left( at+bx\right) \right),
\end{equation*}
where $b^{2}=\frac{\left( k-a\right) }{D}>0$ and $a$ is a constant to be conveniently 
chosen from a known value of $\phi \left( x,t\right)$.

For the fractional mathematical model proposed, we assume that $0<\mu\leq 1$
and the fractional derivative of order $\mu$ is considered in the 
Caputo sense \cite{IP,MF,RM}, defined as follows:

\begin{equation*}
\mathcal{D}_{t}^{\mu }C\left( x,t\right) =\frac{\partial ^{\mu }C\left(
x,t\right) }{\partial t^{\mu }}:=\left\{ 
\begin{array}{c}
\displaystyle\frac{1}{\Gamma \left( n-\mu \right) }\int_{0}^{t}C^{\left( n\right) }\left(
\tau ,t\right) \left( t-\tau \right) ^{n-\mu -1}d\tau ,\text{ \ }n-1<\mu <n
\\ 
\\ 
C^{\left( n\right) }\left( x,t\right),\text{ \ \
\  \ \ \ \ \ \ \ \ \ \ \ \ \ \ \ \ \ \ \ \ \ \ \ \
\ \ \ \ \ \ \ \ \ \ \ \ \ \ }\mu =n, 
\end{array}
\right.
\end{equation*}
where $C^{(n)}(x,t)$ is the usual derivative of order $n$ with respect to $t$.
The particular case $\mu=1$, recovers the result obtained by Sharma et al.\
\cite{GMR}.  Furthermore, we must impose the following initial and boundary
conditions for Eq.(\ref{S1}): 

\begin{equation}
\left\{ 
\begin{array}{ll}
C(x,0)=0, & x\geq 0 \\ 
C(0,t)=1, & t>0 \\ 
C(\infty ,t)=0, & t>0.
\end{array}%
\right.  \label{S4}
\end{equation}

From these considerations, it follows that the time-fractional mathematical model to
be addressed is composed of a non-homogeneous fractional PDE

\begin{equation}  \label{S3}
D_{L}\mathcal{D}_{x}^{2}C\left( x,t\right) - \mathcal{D}_{t}^{\mu }C\left(
x,t\right) =\exp \left( -\left( at+bx\right) \right) , \quad a, b \in \mathbb{R}, 
\end{equation}
with initial and boundary conditions given by Eq.(\ref{S4}).

\section{Analytic Solution}

To solve this problem, we employ the methodology of Laplace
transform to convert the non-homogeneous fractional PDE into a non-homogeneous linear ordinary differential equation. 

Then, applying the Laplace transform \cite{LD,PD} in relation to the time
variable $t$ on both sides of Eq.(\ref{S3}), we have

\begin{equation*}
D_{L}\frac{d^{2}}{dx^{2}}C\left( x,s\right) -s^{\mu }C\left( x,s\right)
+s^{\mu -1}C\left( x,0\right) =\frac{\exp \left( -bx\right) }{s+a}.
\end{equation*}

Using the initial condition $C(x,0)=0$ we can rewrite this equation as 

\begin{equation}  \label{S9}
D_{L}\frac{d^{2}}{dx^{2}}C\left( x,s\right) -s^{\mu }C\left( x,s\right) =%
\frac{\exp \left( -bx\right) }{s+a},
\end{equation}
where $0<\mu \leq 1$, $D_{L}>0$ and
\begin{equation*}
C\left( x,s\right) = \mathscr{L}\left\{ C\left( x,t\right) \right\} =\colon
\int_{0}^{\infty }e^{-st}C\left( x,t\right) dt
\end{equation*}
is the Laplace transform of $C(x,t)$ with parameter $s$, $\mbox{Re}(s)>0$.

Using the methods of characteristic equation and undetermined coefficients in
Eq.(\ref{S9}) we obtain the general solution, given by

\begin{equation}  \label{S15}
C\left( x,s\right) =\left( \frac{1}{s}+\frac{1}{\left( s+a\right) \left(
s^{\mu }-\frac{b^{2}}{\alpha ^{2}}\right) }\right) \exp \left( -\alpha
xs^{\mu /2}\right) +\frac{\exp \left( -bx\right) }{\left( s+a\right) \left( 
\frac{b^{2}}{\alpha ^{2}}-s^{\mu }\right) },
\end{equation}
where $\alpha ^{2}=\frac{1}{D_{L}}$ and $D_{L}>0$.

In order to recover the solution in the time variable we take the inverse
Laplace transform on both sides of Eq.(\ref{S15}), obtaining 

\begin{eqnarray}  \label{S16}
C\left( x,t\right) &=&\mathscr{L}^{-1}\left\{ C\left( x,s\right) \right\} =%
\mathscr{L}^{-1}\left\{ \frac{1}{s}\exp \left( -\alpha xs^{\mu /2}\right)
\right\} +  \notag \\
&&+\mathscr{L}^{-1}\left\{ \frac{1}{\left( s+a\right) \left( s^{\mu }-\frac{%
b^{2}}{\alpha ^{2}}\right) }\exp \left( -\alpha xs^{\mu /2}\right) \right\} 
\notag \\
&&-\mathscr{L}^{-1}\left\{ \frac{1}{\left( s+a\right) \left( s^{\mu }-\frac{%
b^{2}}{\alpha ^{2}}\right) }\exp \left( -bx\right) \right\},
\end{eqnarray}
where

\begin{equation*}
C\left( x,t\right) = \mathscr{L}^{-1}\left\{ C\left( x,s\right) \right\} =\colon \frac{1%
}{2\pi i}\int_{\gamma -i\infty }^{\gamma +i\infty }e^{st}C\left( x,s\right)
ds
\end{equation*}
is the inverse Laplace transform, and the integral is performed in the complex
plane with the singularities $C(x,s)$ on the left side of $\gamma =
\mbox{Re}(s)$.

Introducing the change $\beta ^{2}=b^{2}D_{L}$, we rewrite Eq.(\ref{S16}) as 

\begin{equation*}
C\left( x,t\right) =C_{1}\left( x,t\right) +C_{2}\left( x,t\right)
- \mbox{exp}(-bx) C_{3}\left( x,t\right),
\end{equation*}
with

\begin{eqnarray*}
C_{1}\left( x,t\right)  &=&\mathscr{L}^{-1}\left\{ \frac{\exp \left( -\alpha
xs^{\mu /2}\right) }{s}\right\} ;  \notag \\
C_{2}\left( x,t\right)  &=&\mathscr{L}^{-1}\left\{ \frac{\exp \left( -\alpha
xs^{\mu /2}\right) }{\left( s+a\right) \left( s^{\mu }-\beta ^{2}\right) }%
\right\} ;  \notag \\
C_{3}\left( x,t\right)  &=&\underset{x\rightarrow 0}{\lim }C_{2}\left(
x,t\right) .
\end{eqnarray*}

In order to proceed, we calculate each inverse Laplace transform separately. To
calculate $C_{1}\left(x,t\right)$ we introduce the MacLaurin series associated
with the exponential function; choosing $f^{(k)}(0)=1$ in the series,
we have

\begin{equation}\label{S18}
\frac{1}{s}\exp \left( -\alpha xs^{\mu /2}\right) =\overset{\infty }{%
\underset{k=0}{\sum }}\frac{\left( -\alpha x\right) ^{k}}{k!}s^{\frac{\mu k}{
2}-1}.  
\end{equation}

Applying the inverse Laplace transform on both sides of Eq.(\ref{S18}), we obtain

\begin{equation}\label{S17}
C_{1}\left( x,t\right) =\mathscr{L}^{-1}\left\{ \frac{1}{s}\exp \left(
-\alpha xs^{\mu /2}\right) \right\} =\overset{\infty }{\underset{k=0}{\sum }}
\frac{\left( -\alpha x\right) ^{k}}{k!}\mathscr{L}^{-1}\left\{ s^{\frac{\mu k
}{2}-1}\right\} .
\end{equation}

Using the result

\begin{equation*}  
\mathscr{L}^{-1}\left\{ s^{-q}\right\} =\frac{t^{q-1}}{\Gamma \left( q\right)},
\end{equation*}
with $\mbox{Re}(q)>0$, $q=1-\mu k/2$, we can rewrite Eq.(\ref{S17}) as follows: 

\begin{equation}\label{S22}
C_{1}\left( x,t\right) =\underset{k=0}{\overset{\infty }{\sum }}\frac{\left(
-\alpha x/t^{\mu /2}\right) ^{k}}{k!\Gamma \left( 1-\mu k/2\right) }.
\end{equation}

Moreover, considering $\beta=1$, $\alpha=-\mu/2$ and $z=-\frac{\alpha x}{t^{\mu
/2}}$ in the definition of the Wright function \cite{FG}, we can write

\begin{equation}  \label{S23}
\mathbb{W}\left( -\mu /2,1;z\right) =\overset{\infty }{\underset{k=0}{\sum }}
\frac{z^{k}}{k!\Gamma \left( -\mu k/2+1\right) }.
\end{equation}

Then, from Eq.(\ref{S22}) and Eq.(\ref{S23}) we obtain

\begin{equation}  \label{S24}
C_{1}\left( x,t\right) =\mathbb{W}\left( -\mu /2,1;-\frac{\alpha x}{t^{\mu
/2}}\right) .
\end{equation}

Now we evaluate the second inverse Laplace transform, 

\begin{equation}  \label{S25}
C_{2}\left( x,t\right) =\mathscr{L}^{-1}\left\{ \frac{1}{\left( s+a\right)
\left( s^{\mu }-\beta ^{2}\right) }\exp \left( -\alpha xs^{\mu /2}\right)
\right\} .
\end{equation}

As with $C_{1}(x,t)$, we also write the exponential function in
terms of its MacLaurin series. So, we have

\begin{equation}\label{S26}
\frac{1}{\left( s+a\right) \left( s^{\mu }-\beta ^{2}\right) }\exp \left(
-\alpha xs^{\mu /2}\right) =\overset{\infty }{\underset{m=0}{\sum }}\frac{%
\left( -\alpha x\right) ^{m}}{m!}\frac{s^{\mu m/2}}{\left( s+a\right) \left(
s^{\mu }-\beta ^{2}\right) }.
\end{equation}

Once more, applying the inverse Laplace transform on both sides of Eq.(\ref{S26}), we can write

\begin{equation}  \label{S27}
\mathscr{L}^{-1}\left\{ \frac{1}{\left( s+a\right) \left( s^{\mu }-\beta
^{2}\right) } \exp \left( -\alpha xs^{\mu /2}\right) \right\} =\overset{%
\infty }{\underset{m=0}{\sum }}\frac{\left( -\alpha x\right) ^{m}}{m!}%
\mathscr{L}^{-1}\left\{ \frac{s^{\mu m/2}}{\left( s+a\right) \left( s^{\mu
}-\beta ^{2}\right) }\right\} .
\end{equation}

In order to evaluate this inverse Laplace transform, we consider the following
expression \cite{REJ}: 
\begin{equation*}
\Omega =\frac{s^{\sigma }}{s^{\alpha }+\widetilde{a}s^{\delta }+bs^{\gamma }+cs^{\mu }+d},
\end{equation*}
with $\widetilde{a}, b, c, d\in\mathbb{R}$ and $\alpha, \delta, \gamma, \mu\in\mathbb{R}$ such that $\widetilde{a}\neq 0$ and $\alpha>\delta>\gamma>\mu$.

Assuming the condition $\left\vert \frac{bs^{\gamma }+cs^{\mu }+d}{s^{\alpha
}+\widetilde{a}s^{\delta }}\right\vert <1$ and using the geometric series we
have

\begin{eqnarray}\label{S28}
\overset{\infty }{\underset{k=0}{\sum }}\left( -1\right) ^{k}s^{\sigma
-\delta-\delta k}\frac{\left( bs^{\gamma }+cs^{\mu }+d\right) ^{k}}{\left(
s^{\alpha -\delta }+\widetilde{a}\right) ^{k+1}} &=&\frac{s^{\sigma }}{s^{\alpha
}+s^{\delta }\widetilde{a}}\left( \frac{1}{1+\frac{bs^{\gamma }+cs^{\mu }+d}{s^{\alpha
}+\widetilde{a}s^{\delta }}}\right)   \notag \\
&=&\frac{s^{\sigma }}{bs^{\gamma }+cs^{\mu }+d+s^{\alpha }+\widetilde{a}s^{\delta }}.
\end{eqnarray}

The binomial theorem and the definition of binomial coefficients \cite{JJM}
allow us to write in a convenient way Eq.(\ref{S28}) as 

\begin{eqnarray}\label{S29}
\Omega  &=&\overset{\infty }{\underset{k=0}{\sum }}\left( -1\right) ^{k}%
\overset{k}{\underset{l=0}{\sum }}\binom{k}{l}d^{l}\left( bs^{\gamma
}+cs^{\mu }\right) ^{k-l}\frac{s^{\sigma -\delta -\delta k}}{\left( s^{\alpha
-\delta }+\widetilde{a}\right) ^{k+1}}  \notag \\
&=&\overset{\infty }{\underset{k=0}{\sum }}\left( -1\right) ^{k}\overset{k}{%
\underset{l=0}{\sum }}\frac{k!}{l!\left( k-l\right) !}d^{l}\underset{j=0}{%
\overset{k-l}{\sum }}\frac{\left( k-l\right) !}{j!\left( k-l-j\right) !}%
\left( bs^{\gamma }\right) ^{k-l-j}\left( cs^{\mu }\right) ^{j}\frac{%
s^{\sigma -\delta -\delta k}}{\left( s^{\alpha -\delta }+\widetilde{a}\right) ^{k+1}} 
\notag \\
&=&\overset{\infty }{\underset{k=0}{\sum }}\left( -1\right) ^{k}b^{k}k!%
\overset{k}{\underset{l=0}{\sum }}\frac{\left( d/b\right) ^{l}}{l!}\underset{
j=0}{\overset{k-l}{\sum }}\frac{\left( c/b\right) ^{j}}{j!\left(
k-l-j\right) !}\Lambda _{\sigma },
\end{eqnarray}
where $\Lambda _{\sigma }=\frac{s^{\sigma -\delta \left( 1+k\right) +\mu
j+\gamma\left( k-l-j\right) }}{\left( s^{\alpha -\delta }+\widetilde{a}\right) ^{k+1}}$.

Taking the inverse Laplace transform on both sides of Eq.(\ref{S29}) and using the result

\begin{equation}\label{S31}
\mathscr{L}^{-1}\left\{ \Lambda _{\sigma }\right\} =\mathscr{L}^{-1}\left\{ 
\frac{s^{\sigma -\delta \left( 1+k\right) +\mu j+\gamma \left( k-l-j\right) }%
}{\left( s^{\alpha -\delta }+\widetilde{a}\right) ^{k+1}}\right\} =t^{\xi -1}\mathbb{E}%
_{\alpha -\delta ,\xi }^{k+1}\left( -\widetilde{a}t^{\alpha -\delta }\right) ,
\end{equation}
we get 

\begin{equation}  \label{S32}
\mathscr{L}^{-1}\left\{ \Omega \right\} =\overset{\infty }{\underset{k=0}{%
\sum }}\left( -1\right) ^{k}b^{k}k!\overset{k}{\underset{l=0}{\sum }}\frac{%
\left(d/b\right) ^{l}}{l!}\underset{j=0}{\overset{k-l}{\sum }}\frac{%
\left(c/b\right) ^{j}}{j!\left( k-l-j\right) !}t^{\xi -1} \mathbb{E}_{\alpha
-\delta ,\xi}^{k+1}\left( -\widetilde{a}t^{\alpha -\delta }\right) ,
\end{equation}
with $\xi =-\sigma +\alpha +\left( \alpha -\gamma \right) k+\gamma l-\left(
\mu-\gamma \right) j$ and where $\mathbb{E}_{\alpha -\delta
,\xi}^{k+1}\left(\cdot\right) $ is the three parameters Mittag-Leffler function
\cite{REJ,GKAM}.

In particular, considering $c=0$ in Eq.(\ref{S32}), we have that $j=0$ is the
only term contributing to the sum and we conclude that

\begin{equation}  \label{S33}
\mathscr{L}^{-1}\left\{ \frac{s^{\sigma }}{s^{\alpha }+\widetilde{a}s^{\delta
}+bs^{\gamma }+d}\right\} =\overset{\infty }{\underset{k=0}{\sum }}\left(
-1\right) ^{k}b^{k}k!\overset{k}{\underset{l=0}{\sum }}\frac{\left(
d/b\right) ^{l}t^{\xi -1}}{l!\left( k-l\right) !}\mathbb{E}_{\alpha -\delta
,\xi }^{k+1}\left( -\widetilde{a}t^{\alpha -\delta }\right),
\end{equation}
where $\xi =-\sigma +\alpha +\left( \alpha -\gamma \right) k+\gamma l$ and 
$\alpha >\delta >\gamma $.

Then, putting $\sigma =\mu m/2$, $d=-a\beta ^{2}$, $\alpha =\mu +1$, $\gamma =\mu
$, $\delta =1$, $b=a$ and $\widetilde{a}=-\beta ^{2}$ in Eq.(\ref{S33}) and
going back to Eq.(\ref{S27}), we can write

\begin{equation}\label{S34}
C_{2}\left( x,t\right) =t^{\mu }\underset{m=0}{\overset{\infty }{\sum }}%
\frac{\left( -\alpha xt^{-\mu /2}\right) ^{m}}{m!}\underset{k=0}{\overset{%
\infty }{\sum }}\left( -at\right) ^{k}k!\overset{k}{\underset{l=0}{\sum }}%
\frac{\left( -\beta ^{2}t^{\mu }\right) ^{l}}{l!\left( k-l\right) !}\mathbb{E%
}_{\mu ,\theta }^{k+1}\left( \beta ^{2}t^{\mu }\right) ,
\end{equation}
where $\theta =-\mu m/2+\mu +1+k+\mu l$.

With the aim of obtaining the solution of the PDE in terms of the two-parameters Mittag-Leffler function, we evaluated the sum on $l$
appearing in the last expression in order to find a relationship between
two- and three-parameters Mittag-Leffler functions. Using the
identity

\begin{equation}  \label{S35}
\Lambda =\overset{k}{\underset{j=0}{\sum }}\frac{\left( z\right) ^{j}}{%
j!\left( k-j\right) !}\mathbb{E}_{\lambda ,\lambda j+\delta }^{\rho }\left(
-z\right) =\overset{k}{\underset{j=0}{\sum }}\underset{l=0}{\overset{\infty }%
{\sum }}\frac{\left( z\right) ^{j}}{j!\left( k-j\right) !}\frac{\left( \rho
\right)_{l}\left( -z\right) ^{l}}{l!\Gamma \left( \lambda l+\lambda j+\delta
\right) },
\end{equation}
where $\left( \rho \right) _{l}=\rho \left( \rho +1\right) ....\left(
\rho+l-1\right) $, together with the definition and properties of the binomial
coefficients in Eq.(\ref {S35}), we can write \cite{REJ}

\begin{eqnarray}\label{S40}
\overset{k}{\underset{j=0}{\sum }}\frac{\left( z\right) ^{j}}{j!\left(
k-j\right) !}\mathbb{E}_{\lambda ,\lambda j+\delta }^{\rho }\left( -z\right) 
&=&\underset{i=0}{\overset{\infty }{\sum }}\frac{\left( -z\right) ^{i}}{%
\Gamma \left( \lambda i+\delta \right) }\frac{1}{k!}\overset{k}{\underset{j=0}%
{\sum }}\frac{\left( -1\right) ^{j}k!}{j!\left( k-j\right) !}\binom{i-j+\rho
-1}{\rho -1}  \notag \\
&=&\underset{i=0}{\overset{\infty }{\sum }}\frac{\left( -z\right) ^{i}}{%
\Gamma \left( \lambda i+\delta \right) }\frac{1}{k!}\frac{\left( \rho
-k\right) _{i}}{i!}=\frac{1}{k!}\mathbb{E}_{\lambda ,\delta }^{\rho -k}\left(
-z\right) .
\end{eqnarray}

Choosing $z=-\beta ^{2}t^{\mu }$, $\rho =k+1$, $\lambda =\mu $, $j=l$ and
$\delta=k+\mu +1-\mu m/2$ in Eq.(\ref{S40}) and substituting the result 
into Eq.(\ref{S34}), we conclude that

\begin{equation}  \label{S41}
C_{2}\left( x,t\right) =t^{\mu }\underset{m=0}{\overset{\infty }{\sum }}%
\frac{\left( -\alpha xt^{-\mu /2}\right)^{m} }{m!}\underset{k=0}{\overset{%
\infty }{\sum }}\left( -at\right) ^{k}\mathbb{E}_{\mu ,\mu +k+1-\mu
m/2}\left( \beta ^{2}t^{\mu }\right),
\end{equation}
where $\mathbb{E}_{\alpha ,\beta }\left(\cdot\right) $ is the two-parameters Mittag-Leffler function.

The last inverse Laplace transform, $C_{3}(x,t)$, is obtained by means of a
convenient limit, i.e., we consider $x\rightarrow 0 $ in the Eq.(\ref{S41}).
The only term that contributes in this limit is $m=0$, i.e., we get

\begin{equation}\label{S42}
C_{3}\left( x,t\right) =t^{\mu }\underset{k=0}{\overset{\infty }{\sum }}%
\left( -at\right) ^{k}\mathbb{E}_{\mu ,\mu +k+1}\left( \beta ^{2}t^{\mu
}\right) .
\end{equation}

Thus, from the results obtained in Eq.(\ref{S24}), Eq.(\ref{S41}) and
Eq.(\ref{S42}), we get the solution associated with our initial problem, i.e.,
a solution of Eq.(\ref{S3}) satisfying the conditions given by Eq.(\ref{S4})

\begin{eqnarray}  \label{S43}
C\left( x,t\right) &=&t^{\mu }\underset{m=0}{\overset{\infty }{\sum }}\frac{%
\left( -\alpha xt^{-\mu /2}\right) ^{m}}{m!}\underset{k=0}{\overset{\infty }{%
\sum }}\left( -at\right) ^{k}\mathbb{E}_{\mu ,\mu +k+1-\mu m/2}\left( \beta
^{2}t^{\mu }\right) +  \notag \\
&&+\mathbb{W}\left( -\mu /2,1;-\frac{\alpha x}{t^{\mu /2}}\right) -\exp
\left( -bx\right) t^{\mu }\underset{k=0}{\overset{\infty }{\sum }}\left(
-at\right) ^{k}\mathbb{E}_{\mu ,\mu +k+1}\left( \beta ^{2}t^{\mu }\right) ,
\end{eqnarray}
where the parameters are given by $\alpha ^{2}=1/D_{L}$, $\beta ^{2}=b^{2}D_{L}$ and $0<\mu \leq 1$. 

Let us now perform a graphical analysis. For this sake, we have
to choose values for some parameters appearing in the solution given by
Eq.(\ref{S43}). We used the following values: axial dispersion coefficient
$D_{L}= 4.8\times 10^{-4}\mbox{cm}^2 \mbox{s}^{-1}$ \cite{WHRJ}; diffusivity
coefficient of oxygen $D=9.8\times 10^{-5}\mbox{cm}^2 \mbox{s}^{-1}$
\cite{JRP}; nutrient transfer coefficient $k=1.5\times
10^{-4}\mbox{m}\mbox{s}^{-1}$ \cite{AEJ}; $a=-0.005\times
10^{-4}\mbox{m}\mbox{s}^{-1}$ \cite{AEJ}.  We also fix a
time $t=15\mbox{s}$ and we consider a certain interval $x = [0,4]$, which can be
extended.

\begin{figure}[!ht]
\caption{Analytic solution of fractional order PDE, Eq.(\ref{S43}).}
\centering 
\includegraphics[width=10cm]{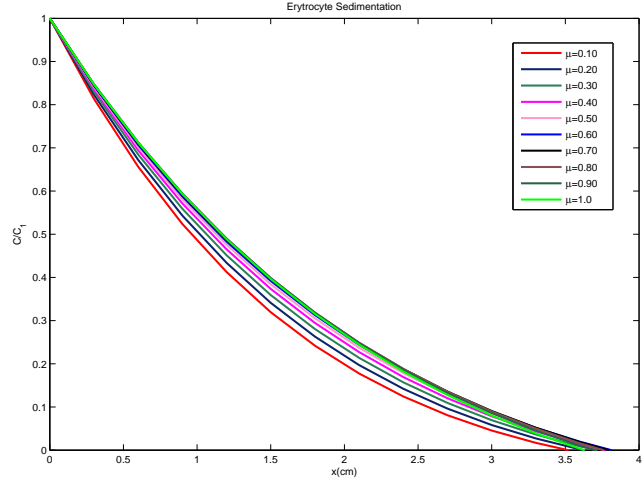} 
\label{fig:eryt}
\end{figure}

\begin{figure}[!ht]
\caption{Analytic solution of entire order PDE.}
\centering 
\includegraphics[width=10cm]{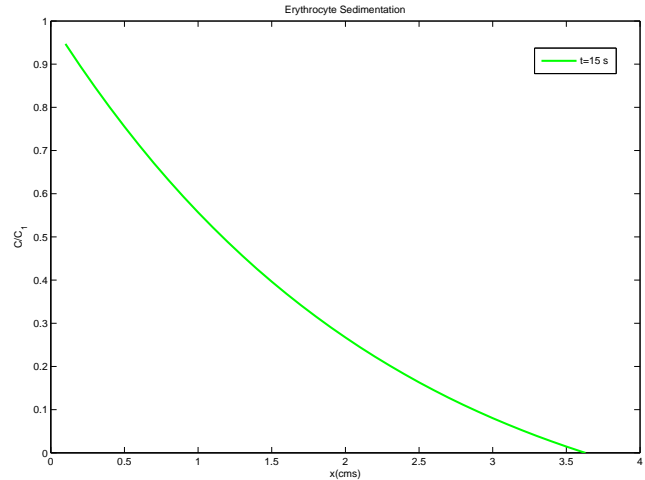} 
\label{fig:eryt1}
\end{figure}

\newpage
In figures 1 and 2, the horizontal axis $x$ represents space and the vertical
axis $y$ is the normalized concentration of nutrients in blood.

The parameter values used to plot figure 1 were also used to plot the solution
of the integer order PDE, figure 2.  The graphics were plotted using MATLAB
7:10 software (R2010a).

Remark that as $x$ (space) increases, the value of $C/C_{1}$ (concentration of nutrients) decreases, that is, when we move towards the extremity of the artery ($x$=0), the blood concentration of solute decreases. A decrease in solute concentration means that cells are not
enough efficient in getting their nutrition, so we conclude that the efficiency
of nutrient transport near the artery is greater than at its venous extremity.

With the freedom given to parameter $\mu$ ($0<\mu\leq1$), it is possible to
describe more accurately the information about the concentration of nutrients
near the arterial extremity because, as seen above the fractionalization of the
derivative refines the solution. Note that for $\mu = 0.10 $ the behavior of the analytic solution remains near the arterial (x = 0) for longer time. We can thus see that as $\mu\rightarrow  1$, the fractional solution converges to the solution of the integer order PDE.

We supposed that the space variable $x$ lies within the range $[0,4]$. We might as well examine variable $x$ in the range $[0,12]$ or any other interval; however the first representative feature is that because for $x\geq3.8$ the level $C/C_{1}$ remains below the $x$ axis. So it interesting, in this context, to do analysis only on the $[0,4]$ range.
 

\section{Particular Case: $\mu \rightarrow 1$}

In this section, we analyze the solution of the fractional PDE in the limit $\mu \rightarrow 1 $, in order to recover the result found by Sharma et al.\ \cite{GMR}.

Since the solution of the fractional PDE Eq.(\ref{S3}) is given by Eq.(\ref{S43}), taking the limit 
$\mu\rightarrow 1 $, it follows that

\begin{eqnarray}  \label{S44}
C\left( x,t\right) &=&t\underset{m=0}{\overset{\infty }{\sum }}\frac{\left(
-\alpha xt^{-1/2}\right) ^{m}}{m!}\underset{k=0}{\overset{\infty }{\sum }}%
\left( -at\right) ^{k}\mathbb{E}_{1,k+2-m/2}\left( \beta ^{2}t\right)+ 
\notag \\
&&+\mathbb{W}\left( -1/2,1;-\frac{\alpha x}{t^{1/2}}\right) -\exp \left(
-bx\right) t\underset{k=0}{\overset{\infty }{\sum }}\left( -at\right)^{k}%
\mathbb{E}_{1,k+2}\left( \beta ^{2}t\right) .
\end{eqnarray}

In the last two terms of the sum in Eq.(\ref{S44}), we can use the following
results \cite{MA}: 

\begin{equation*}  
\mathbb{W}\left( -1/2,1;-x\right) = \mbox{erf}c\left( x/2\right) 
\end{equation*}
and

\begin{equation} \label{S46}
t\overset{\infty }{\underset{k=0}{\sum }}\left( -at\right) ^{k}\mathbb{E}%
_{1,2+k}\left( \beta ^{2}t\right) =\frac{\exp \left( \beta ^{2}t\right)
-\exp \left( -at\right) }{a+\beta ^{2}},
\end{equation}
Thus, Eq.(\ref{S44}) can be written as 

\begin{eqnarray}  \label{S47}
C\left( x,t\right) &=&t\underset{m=0}{\overset{\infty }{\sum }}\frac{\left(
-\alpha xt^{-1/2}\right) ^{m}}{m!}\underset{k=0}{\overset{\infty }{\sum }}%
\left( -at\right) ^{k}\mathbb{E}_{1,k+2-m/2}\left( \beta ^{2}t\right) + 
\notag \\
&&+1+\mbox{erf}\left( -\alpha x/2\sqrt{t}\right) -\exp \left( -bx\right) 
\frac{\exp \left( \beta ^{2}t\right) -\exp \left( -at\right) }{a+\beta ^{2}}.
\end{eqnarray}

To recover the result presented by Sharma et al.\ \cite{GMR}, we need to express
Eq.(\ref{S47}) in terms of $\mbox{erfc}(\cdot)$ and $\mbox{exp}(\cdot)$
functions, since their solution of the convection-diffusion equation is given
in terms of the product of $\mbox{exp}(\cdot)$ and $\mbox{erfc}(\cdot)$
functions. With this aim, we evaluate the inverse Laplace transform in
Eq.(\ref{S25}) using partial fractions.

Taking the limit $\mu \rightarrow 1$ in Eq.(\ref{S25}), it follows that

\begin{equation}  \label{S48}
C_{2}\left( x,t\right) =\mathscr{L}^{-1}\left\{ \frac{\exp \left( -\alpha x%
\sqrt{s}\right) }{\left( s+a\right) \left( s-\beta ^{2}\right) }\right\} .
\end{equation}

Using partial fractions, we rewrite $G_{a}^{\beta }\left( s\right)
=\frac{1}{\left( s+a\right) \left( s-\beta ^{2}\right) }$ as follows:  
        
\begin{eqnarray}  \label{S50}
\frac{2\left( \beta ^{2}+a\right) }{\left( s+a\right) \left( s-\beta
^{2}\right) } &=&-\frac{1}{\sqrt{s}\left( \sqrt{s}-i\sqrt{a}\right) }-\frac{1%
}{\sqrt{s}\left( \sqrt{s}+i\sqrt{a}\right) }+  \notag \\
&&+\frac{1}{\sqrt{s}\left( \sqrt{s}-\beta \right) }+\frac{1}{\sqrt{s}\left( 
\sqrt{s}+\beta \right) }.
\end{eqnarray}

Multiplying Eq.(\ref{S50}) by $\exp \left( -\alpha x\sqrt{s}\right)$ and taking the inverse Laplace transform of both sides, we have

\begin{eqnarray}  \label{S51}
2\left( \beta ^{2}+a\right) \mathscr{L}^{-1}\left\{ \frac{\exp \left(
-\alpha x\sqrt{s}\right) }{\left( s+a\right) \left( s-\beta ^{2}\right) }%
\right\} &=&-\mathscr{L}^{-1}\left\{ \frac{\exp \left( -\alpha x\sqrt{s}%
\right) }{\sqrt{s}\left( \sqrt{s}-i\sqrt{a}\right) }\right\} -\mathscr{L}%
^{-1}\left\{ \frac{\exp \left( -\alpha x\sqrt{s}\right) }{\sqrt{s}\left( 
\sqrt{s}+i\sqrt{a}\right) }\right\} +  \notag \\
&&+\mathscr{L}^{-1}\left\{ \frac{\exp \left( -\alpha x\sqrt{s}\right) }{%
\sqrt{s}\left( \sqrt{s}-\beta \right) }\right\} +\mathscr{L}^{-1}\left\{ 
\frac{\exp \left( -\alpha x\sqrt{s}\right) }{\sqrt{s}\left( \sqrt{s}+\beta
\right) }\right\} .
\end{eqnarray}

In evaluating the inverse Laplace transforms, we use the following result
\cite{FEJ}:

\begin{equation}  \label{S52}
\mathscr{L}^{-1}\left\{ \frac{\exp \left( -k\sqrt{s}\right) }{\sqrt{s}\left( 
\sqrt{s}+b\right) }\right\} =\exp \left( bk\right) \exp \left( b^{2}t\right) %
\mbox{erfc}\left( b\sqrt{t}+\frac{k}{2\sqrt{t}}\right) ,
\end{equation}
with $k\geq 0$, $b\in \mathbb{C}$ and 
$\mbox{erfc}\left( x\right) $ the complementary error function.

Thus, applying Eq.(\ref{S52}) in each term of Eq.(\ref{S51}), we have

\begin{eqnarray}  \label{S53}
2\left( \beta ^{2}+a\right) \mathscr{L}^{-1}\left\{ \frac{\exp \left(
-\alpha x\sqrt{s}\right) }{\left( s+a\right) \left( s-\beta ^{2}\right) }%
\right\} &=&\exp \left( \beta \alpha x\right) \exp \left( \beta ^{2}t\right) %
\mbox{erfc}\left( \beta \sqrt{t}+\frac{\alpha x}{2\sqrt{t}}\right) \\
&&+\exp \left( -\beta \alpha x\right) \exp \left( \beta ^{2}t\right) %
\mbox{erfc}\left( -\beta \sqrt{t}+\frac{\alpha x}{2\sqrt{t}}\right)  \notag
\\
&&-\exp \left( i\alpha \sqrt{a}x\right) \exp \left( -at\right) \mbox{erfc}%
\left( i\sqrt{at}+\frac{\alpha x}{2\sqrt{t}}\right)  \notag \\
&&-\exp \left( -i\alpha \sqrt{a}x\right) \exp \left( -at\right) \mbox{erfc}%
\left( -i\sqrt{at}+\frac{\alpha x}{2\sqrt{t}}\right) .  \notag
\end{eqnarray}

Moreover, considering the same particular case $x=0$ in Eq.(\ref{S53}) we
find the same result found for the inverse Laplace transform $C_ {3}(x,t)$,
i.e.,

\begin{eqnarray}\label{S54}
2\left( \beta ^{2}+a\right) \mathscr{L}^{-1}\left\{ \frac{1}{\left(
s+a\right) \left( s-\beta ^{2}\right) }\right\}  &=&\exp \left( \beta
^{2}t\right) \left( \mbox{erfc}\left( \beta \sqrt{t}\right) +\mbox{erfc}\left( -\beta \sqrt{t}\right) \right) -  \notag \\
&&-\exp \left( -at\right) \left( \mbox{erfc}\left( i\sqrt{at}\right) +%
\mbox{erfc}\left( -i\sqrt{at}\right) \right) .
\end{eqnarray}

Analyzing the error functions in Eq.(\ref{S54}), we conclude that

\begin{equation}  \label{S55}
\mathscr{L}^{-1}\left\{ \frac{1}{\left( s+a\right) \left( s-\beta
^{2}\right) }\right\} =\frac{\exp \left( \beta ^{2}t\right) -\exp \left(
-at\right) }{\beta ^{2}+a}.
\end{equation}

As we evaluated the inverse Laplace transform of $C_ {2}(x,t)$ in
Eq.(\ref{S25}) in the case $\mu = 1$ using two different procedures, i.e.
Eq.(\ref{S44}) and Eq.(\ref{S53}), involving respectively a Mittag-Leffler
function and error functions, we can write the interesting identity

\begin{eqnarray}  \label{S56}
&&2\left( a+\beta ^{2}\right) t\underset{m=0}{\overset{\infty }{\sum }}\frac{%
\left( -\alpha xt^{-1/2}\right) }{m!}\overset{\infty }{\underset{k=0}{\sum }}%
\left( -at\right) ^{k}\mathbb{E}_{1,2+k-m/2}\left( \beta ^{2}t\right) 
\begin{array}{c}
=
\end{array}
\notag \\
&=&e^{\beta \alpha x}e^{\beta ^{2}t}\mbox{erfc}\left( \beta \sqrt{t}+\frac{%
\alpha x}{2\sqrt{t}}\right) +e^{-\beta \alpha x}e^{\beta ^{2}t}\mbox{erfc}%
\left( -\beta \sqrt{t}+\frac{\alpha x}{2\sqrt{t}}\right) -  \notag \\
&&-e^{i\alpha \sqrt{a}x}e^{-at}\mbox{erfc}\left( i\sqrt{at}+\frac{\alpha x}{2%
\sqrt{t}}\right) -e^{-i\alpha \sqrt{a}x}e^{-at}\mbox{erfc}\left( -i\sqrt{at}+%
\frac{\alpha x}{2\sqrt{t}}\right).
\end{eqnarray}

Further, considering $\alpha=0$ in Eq.(\ref{S56}), which means that only $m=0$
contributes to the first sum, we obtain

\begin{eqnarray*}
2\left( a+\beta ^{2}\right) t\overset{\infty }{\underset{k=0}{\sum }}%
\left(-at\right) ^{k}\mathbb{E}_{1,2+k}\left( \beta ^{2}t\right) &=&2\left(
e^{\beta ^{2}t}\mbox{erfc}\left( \beta \sqrt{t}\right) +e^{\beta ^{2}t}%
\mbox{erfc}\left(-\beta \sqrt{t}\right) \right) - \\
&&-2\left( e^{-at}\mbox{erfc}\left( i\sqrt{at}\right) -e^{-at}\mbox{erfc}%
\left( -i\sqrt{at}\right) \right) \\
&=&2\left( e^{\beta ^{2}t}-e^{-at}\right) ,
\end{eqnarray*}
or, rearranging, 

\begin{equation*}
t\overset{\infty }{\underset{k=0}{\sum }}\left( -at\right) ^{k}\mathbb{E}%
_{1,2+k}\left( \beta ^{2}t\right) =\frac{e^{\beta ^{2}t}-e^{-at}}{\left(
a+\beta ^{2}\right) },
\end{equation*}
which, in fact, justifies the result given by Eq.(\ref{S46}). Consequently, Eq.(\ref{S56}) can be interpreted as a generalization of Eq.(\ref{S46}). 
Also, considering $a=0$ in the previous equation, we have

\begin{equation*}
\beta ^{2}t\mathbb{E}_{1,2}\left( \beta ^{2}t\right) =e^{\beta ^{2}t}-1,
\end{equation*}
which is an identity involving the Mittag-Leffler function \cite{GKAM}.

Finally, as we calculated the inverse Laplace transform in the case $\mu=1$ by
two different ways, Eq.(\ref{S44}) and Eq.(\ref {S53}), we can write the main
relation we need to recover the solution proposed by Sharma et al.\ \cite{GMR},
that is, by Eq.(\ref{S47}) and Eq.(\ref{S56}): 

\begin{eqnarray}\label{S57}
C\left( x,t\right)  &=&t\underset{m=0}{\overset{\infty }{\sum }}\frac{\left(
-\alpha xt^{-1/2}\right) ^{m}}{m!}\underset{k=0}{\overset{\infty }{\sum }}%
\left( -at\right) ^{k}\mathbb{E}_{1,k+2-m/2}\left( \beta ^{2}t\right) + 
\notag \\
&&+1+\mbox{erf}\left( -\alpha x/2\sqrt{t}\right) -\exp \left( -bx\right) 
\frac{\exp \left( \beta ^{2}t\right) -\exp \left( -at\right) }{a+\beta ^{2}}
\notag \\
&=&1-\mbox{erf}\left( \alpha x/2\sqrt{t}\right) -\frac{\exp \left(
-bx\right) }{a+\beta ^{2}}\left( \exp \left( \beta ^{2}t\right) -\exp \left(
-at\right) \right)   \notag \\
&&+\frac{\exp \left( \beta ^{2}t\right) }{2\left( a+\beta ^{2}\right) }\left[
\begin{array}{c}
\exp \left( \beta \alpha x\right) \mbox{erfc}\left( \beta \sqrt{t}+\frac{%
\alpha x}{2\sqrt{t}}\right)  \\ 
+\exp \left( -\beta \alpha x\right) \mbox{erfc}\left( -\beta \sqrt{t}+\frac{%
\alpha x}{2\sqrt{t}}\right) 
\end{array}%
\right] -  \notag \\
&&-\frac{\exp \left( -at\right) }{2\left( a+\beta ^{2}\right) }\left[ 
\begin{array}{c}
\exp \left( i\alpha \sqrt{a}x\right) \mbox{erfc}\left( i\sqrt{at}+\frac{%
\alpha x}{2\sqrt{t}}\right)  \\ 
+\exp \left( -i\alpha \sqrt{a}x\right) \mbox{erfc}\left( -i\sqrt{at}+\frac{%
\alpha x}{2\sqrt{t}}\right) 
\end{array}%
\right] .
\end{eqnarray}

We emphasize that parameters $D$ and $D_{L}$ are positive constants and
$k\geq a$, as conveniently imposed in both models, the one proposed by Sharma et al.
\cite{GMR} and our fractional version, discussed in Section 2. Moreover,
returning to the original parameters $\beta =\sqrt{\frac{\left( k-a\right)
}{D}D_{L}}$, $b=\sqrt{\frac{k-a}{D}}$, from Eq.(\ref{S57}), we conclude that

\begin{eqnarray}\label{S58}
C\left( x,t\right)  &=&\mbox{erf}c\left( \frac{x}{2\sqrt{D_{L}t}}\right) -%
\frac{\exp \left( -\sqrt{\frac{k-a}{D}}x\right) }{k\left( \frac{D_{L}}{D}%
\right) +a\left( 1-\frac{D_{L}}{D}\right) }\left( \exp \left( \left( \frac{%
k-a}{D}\right) D_{L}t\right) -\exp \left( -at\right) \right) +  \notag \\
&&+\frac{\exp \left( \left( \frac{k-a}{D}\right) D_{L}t\right) }{2\left[
k\left( \frac{D_{L}}{D}\right) +a\left( 1-\frac{D_{L}}{D}\right) \right] }%
\left[ 
\begin{array}{c}
\exp \left( \sqrt{\frac{k-a}{D}}x\right) \mbox{erfc}\left( \frac{x+2D_{L}t%
\sqrt{\frac{k-a}{D}}}{2\sqrt{D_{L}t}}\right)  \\ 
+\exp \left( -\sqrt{\frac{k-a}{D}}x\right) \mbox{erfc}\left( \frac{x-2D_{L}t%
\sqrt{\frac{k-a}{D}}}{2\sqrt{D_{L}t}}\right) 
\end{array}%
\right] -  \notag \\
&&-\frac{\exp \left( -at\right) }{2\left[ k\left( \frac{D_{L}}{D}\right)
+a\left( 1-\frac{D_{L}}{D}\right) \right] }\left[ 
\begin{array}{c}
\displaystyle\exp \left( \frac{i\sqrt{a}x}{\sqrt{D_{L}}}\right) \mbox{erfc}\left( \frac{%
x+2it\sqrt{D_{L}a}}{2\sqrt{D_{L}t}}\right)  \\ 
+\exp \left( -\frac{i\sqrt{a}x}{\sqrt{D_{L}}}\right) \mbox{erfc}\left( \frac{%
x-2it\sqrt{D_{L}a}}{2\sqrt{D_{L}t}}\right) 
\end{array}%
\right] ,
\end{eqnarray}
which is exactly the result obtained in \cite{GMR}.

\section{Concluding Remarks} 

After a brief introduction to the study of the concentration of nutrients in
blood, a factor that interferes with ESR, by means of a fractional mathematical
model employing fractional derivatives in the Caputo sense, we obtained its
analytic solution in terms of the Mittag-Leffler function and the Wright
function using the methodology of Laplace transform in the time variable.  We
should point out that one of the greatest challenges of fractional calculus, in
the study of differential equations, is to propose a fractional differential
equation whose corresponding analytic solution recovers the integer order case
in a convenient limit. Here, it was possible to recover the solution of the
integer case applying the limit $\mu\rightarrow 1$ to the analytic solution,
Eq.(\ref{S43}), of the fractional PDE, Eq.(\ref{S3}).
As for what was expected about the relation between the fractional mathematical
model and the integer order model of \cite{GMR}, we can say that our fractional
model provides more accurate information about the concentration of nutrients
in blood.

A natural continuation of this work is to confront our fractional model with
laboratory data, in order to be able to make predictions using the ESR test.
Studies in this direction are being done \cite{JLEC} and will be published in
the near future.


\section*{Acknowledgment}

We thank Prof. Dr. Felix Silva Costa and Dr. J. Em\'{\i}lio Maiorino by the fruitful discussions and Prof. Dr.
F. Mainardi for suggesting several references on the subject.

\end{document}